\documentclass[12pt]{article}
\usepackage[margin=0.7in]{geometry}

\usepackage{color}
\usepackage{multirow}
\usepackage{graphicx}
\usepackage{dcolumn}
\usepackage{bm}
\usepackage{amsfonts}
\usepackage{sidecap}
\usepackage{float}
\usepackage{parskip}
\usepackage{grffile}

\usepackage{color}
\usepackage{hyperref}
\usepackage{hhline}
\usepackage{mathtools}
\usepackage[normalem]{ulem}
\DeclarePairedDelimiter{\norm}{\lVert}{\rVert}
\usepackage{graphicx}
\usepackage{dcolumn}
\usepackage{bm}
\usepackage{hyperref}
\usepackage{sidecap}
\usepackage{float}
\usepackage{parskip}
\usepackage{grffile}
\usepackage{color}
\usepackage{hhline}
\usepackage{xcolor}
\usepackage{adjustbox}
\usepackage{algorithm}
\usepackage{algpseudocode}

\usepackage{hyperref}
\usepackage{scalerel}
\usepackage{tikz}
\usetikzlibrary{svg.path}
\definecolor{orcidlogocol}{HTML}{A6CE39}
\tikzset{
  orcidlogo/.pic={
    \fill[orcidlogocol] svg{M256,128c0,70.7-57.3,128-128,128C57.3,256,0,198.7,0,128C0,57.3,57.3,0,128,0C198.7,0,256,57.3,256,128z};
    \fill[white] svg{M86.3,186.2H70.9V79.1h15.4v48.4V186.2z}
                 svg{M108.9,79.1h41.6c39.6,0,57,28.3,57,53.6c0,27.5-21.5,53.6-56.8,53.6h-41.8V79.1z M124.3,172.4h24.5c34.9,0,42.9-26.5,42.9-39.7c0-21.5-13.7-39.7-43.7-39.7h-23.7V172.4z}
                 svg{M88.7,56.8c0,5.5-4.5,10.1-10.1,10.1c-5.6,0-10.1-4.6-10.1-10.1c0-5.6,4.5-10.1,10.1-10.1C84.2,46.7,88.7,51.3,88.7,56.8z};
  }
}

\newcommand\orcidicon[1]{\href{https://orcid.org/#1}{\mbox{\scalerel*{
\begin{tikzpicture}[yscale=-1,transform shape]
\pic{orcidlogo};
\end{tikzpicture}
}{|}}}}
\begin{document}

\begin{flushleft}
{\large
\textbf{Identification of Chimera using Machine Learning}
}
\\
M.A. Ganaie \orcidicon{0000-0002-3986-4434} $^{\dagger,1}$, Saptarshi Ghosh \orcidicon{0000-0001-8861-5269} $^{\dagger,2}$, Naveen Mendola$^{2}$, M Tanveer \orcidicon{0000-0002-5727-3697}  $^{1,\star}$ \& Sarika Jalan \orcidicon{0000-0002-9984-6107} $^{2,3,\ast}$\\

\it $^{1}$ Discipline of Mathematics, Indian Institute of Technology Indore, Khandwa Road, Simrol, 453552 Indore India\\
\it ${^2}$ Complex Systems Lab, Discipline of Physics, Indian Institute of Technology Indore, Indore 452017, India\\
\it ${^3}$ Discipline of Biosciences and Biomedical Engineering, Indian Institute of Technology Indore, Indore 452017, India\\

$\ast$ E-mail: sarika@iiti.ac.in (Corresponding Author) \\
$\star$ E-mail: mtanveer@iiti.ac.in \\
$^\dagger$ Contributed Equally\\

\date{\today}
\end{flushleft}

\begin{abstract}
Chimera state refers to coexistence of coherent and non-coherent phases in identically coupled dynamical units found in various complex dynamical systems. Identification of Chimera, on one hand is essential due to its applicability in various areas including neuroscience, and on other hand is challenging due to its widely varied appearance in different systems and the peculiar nature of its profile. Therefore, a simple yet universal method for its identification remains an open problem. Here, we present a very distinctive approach using machine learning techniques to characterize different dynamical phases and identify the chimera state from given spatial profiles generated using various different models.
The experimental results show that the performance of the classification algorithms varies for different dynamical models.
The machine learning algorithms, namely random forest, oblique random forest based on tikhonov, parallel-axis split and null space regularization achieved more than $96\% $ accuracy for the Kuramoto model. For the logistic-maps, random forest and tikhonov regularization based oblique random forest showed more than $90\%$ accuracy, and for the H\'enon-Map model, random forest, null-space and axis-parallel split regularization based oblique random forest achieved more than $80\%$ accuracy.
The oblique random forest with null space regularization achieved consistent performance (more than $83\%$ accuracy) across different dynamical models while the auto-encoder based random vector functional link neural network showed relatively lower performance.
This work provides a direction for employing machine learning techniques to identify dynamical patterns arising in coupled non-linear units on large-scale, and for characterizing complex spatio-temporal patterns in real-world systems for various applications.
  \end{abstract}
  
\section*{}
{\bf Studies of coupled dynamics on complex networks have been enormously helpful in getting insight into various dynamical patterns that appear in a wide variety of large-scale real-world complex systems. Chimera state is one of such exotic patterns arising in identically coupled oscillators (for example, in regular networks) which has drawn tremendous attention due to its strangeness and new-found applicability. However, various identification techniques reported in literature are as widely varied as various emergent chimera profiles in different systems. Here, we report a simple yet universal approach to this challenging problem of identification of Chimera by using machine learning. Recently, machine learning (ML) techniques have demonstrated tremendous potential in various areas of nonlinear dynamics and network science. We present a universal machine learning-based approach for the identification of chimera states by utilizing five ML algorithms, namely random forest, oblique random forests via multi-surface proximal support vector machines with tikhonov regularization, axis-parallel split regularization and null Space regularization (MPRaF-T, P, N) and sparse pre-trained / auto-encoder based random vector functional link neural network (RVFL-AE), trained on mere dynamical spatial configurations as input. These techniques can characterize different dynamical states and identify chimera patterns in various time-discrete and time-continuous model systems.}

\section{Introduction}
In 2002, Kuramoto and Battogtokh~\cite{kuramoto2002coexistence} reported a strange co-existence of phase locked and drifting oscillators in non-locally coupled network of identical oscillators. Later on, Abrams and Strogatz christened the state as the chimera state after a fire-breathing hybrid monster from Greek mythology and provided a mathematical foundation to its appearance~\cite{abrams2004chimera}. Since then, the chimera state has been reported to occur in numerous complex systems ranging from network of opto-electronic oscillators~\cite{hart2016experimental} to brain~\cite{bansal2019cognitive,Chouzouris2018} leading to a wide deviation from the initial definition and stringent conditions proposed by Kuramoto {\it et al.} and Abrams {\it et al.} in their landmark papers. Its exotic nature and appearance in a wide variety of real-world complex systems led to an explosive interest in chimera yielding a huge literature in theoretical~\cite{ghosh2016emergence,omelchenko2011loss,bera2016chimera,ujjwal2013chimeras,meena2016chimera,strelkova2018synchronization,semenova2015does} as well as experimental~\cite{kumar2017partially,lazarides2015chimeras,martens2013chimera,nkomo2013chimera} investigations. However, the chimera state reported for a diverse range of systems manifest into diverse forms in terms of variations in shape, appearances, mobility, spatio-temporal behavior and many more~\cite{panaggio2015chimera,scholl2016synchronization}. Chimeras has been dubbed as spatio-temporal phenomenon mostly in time-continuous systems~\cite{kuramoto2002coexistence,abrams2004chimera} whereas time-discrete maps primarily exhibit chimera state with only spatial chaos with simple temporal behavior (mostly periodic)~\cite{ghosh2016emergence,omelchenko2011loss}. A third type of chimera state has also been reported with both temporal as well as spatial chaotic behavior~\cite{dudkowski2014different}. For example, an amplitude chimera corresponds to the coexistence of coherent and incoherent behavior in amplitude with all oscillators possessing the same average frequency~\cite{zakharova2014chimera}. Breathing chimera states correspond to periodic appearance of the state, whereas in travelling chimera state, the incoherent part is non-static in time\cite{abrams2008solvable,xie2014multicluster,schmidt2015chimeras}. 
In addition to previously reported chimera in 1D (regular) networks, chimera state has been reported for 2D~\cite{laing2009dynamics,Schmidt2017,Hizanidis2020} and 3D~\cite{maistrenko2015chimera,Kasimatis2018}, systems as well. Additionally, both purely local~\cite{laing2015chimeras,bera2016chimera} and global coupling~\cite{sethia2014chimera} schemes have been shown to yield a chimera state. 

Overall, the field of chimera has been related to different areas in past-few years. Due to this disparate expansion, the identification and classification of a Chimera state have become a huge challenge. A plethora of measures have been proposed towards the same goal. To describe a few of the proposed measures are as follows, Kemeth {\it et al.} reported a generalized Laplacian correlation measure to classify different types of chimera states~\cite{kemeth2016classification}; Gopal {\it et al.} presented a measure based on strength of incoherence and the discontinuity to identify single and multi-chimera state for the underlying parameter space~\cite{gopal2014observation}; a local order parameter based measure has been widely used to identify chimera states in phase oscillators~\cite{wolfrum2011spectral,omelchenko2013nonlocal}; Hizanidis {\it et al.} presented a chimera-like and metastability indices to recognize chimera states~\cite{hizanidis2016chimera}. However, recent surge in finding new applications of chimera state calls for a universal method for its identification without imposing any numerical restrictions as well as should be applicable regardless of the underlying complex nature of the considered model.

This article embarks on a distinct approach to propose an efficient method to identify chimera states using machine learning technique. Significant advances have been made in the field of machine learning, leading to its extensive usage in diverse fields ranging from astrophysics to natural language recognition, image processing to bio-medical applications~\cite{Mitchell2017,Burkov2019}. Recently, a series of articles related to the implementation of machine learning techniques into non-linear dynamics, has been added to the literature. The recent surge in exploiting machine learning techniques in the investigation of complex dynamical behavior from {\it reconstruction of attractor of chaotic dynamical systems}~\cite{Lu2018,Pathak2017} or {\it hybrid forecasting of chaotic processes}~\cite{Pathak2018} to {\it predicting dynamical observable from network structure}~\cite{Rodrigues2019}, has initiated a new approach to address well-known research problems in nonlinear dynamics. For example, the machine learning techniques have been used to analyse complex spatio-temporal patterns in coupled dynamical systems~\cite{Neofotistos2019,Hart2019,Barmparis2020}.

This article uses multiple machine learning algorithms to identify the chimera states. Based on the recent survey \cite{fernandez2014we}, random forest (RaF) \cite{breiman2001random}  emerged as the best classifier. Therefore, we used Random forest and its ensembles \cite{zhang2014oblique} as the classification algorithms. The ensembles of the RaF (MPRaF-T, P, N) are based on the different regularization techniques used in the multisurface proximal support vector machine (MPSVM). At each node of the decision trees, MPSVM generates the oblique split hyperplane based on the multiple features of  the data. Unlike RaF, MPSVM based oblique random forests  capture the geometric structure of the data and hence show a better performance \cite{zhang2014oblique, TBRaF}. Moreover, we also use the sparse pre-trained functional link network  \cite{zhang2019unsupervised}, also known as auto-encoder based random vector functional link network (RVFL-AE), to learn an efficient feature representation of the data based on the $l_1$ norm regularized autoencoder. The autoencoder explores the hidden feature information which helps in the better performance \cite{zhang2019unsupervised}. 

We employ five advanced machine learning (ML) algorithms, namely RaF, MPRaF-T, MPRaF-P, MPRaF-N and RVFL-AE, to identify chimera states from underlying spatial data. We argue that, due to the generalized approach and adaptive nature of the methods, they can be easily applied to find co-existing hybrid patterns of Chimera state for given set of parameters ranging from amplitude or frequency data of phase oscillators to electrode data from EEG readings. With the help of four different coupled dynamics namely, Kuramoto oscillators and FitzHugh–Nagumo oscillators for time-continuous as well as coupled Logistic maps and He\`non maps for time-discrete dynamics on network of identical couplings i.e. regular networks, we validate ML algorithms in recognizing the complex spatio-temporal patterns. Our approach demonstrate high accuracy rate in categorizing them in Chimera, coherent and incoherent states.  Table \ref{tab:test_acc} depicts that most of the machine learning algorithms show  competitive performance for identifying the chimera states. One can see that all the machine learning algorithms RaF, MPRaF-T, MPRaF-P, MPRaF-N and RVFL-AE show more than $93\% $ accuracy in the coupled Kuramoto oscillators model. In the coupled Logistic maps model, RaF and MPRaF-T showed more than  $90\%$ accuracy while as in the coupled H\'enon maps model MPRaF-N, MPRaF-P and RaF achieved more than $80\%$ accuracy. In the coupled FHN model, MPRaF-N showed relatively better performance as compared to other classification algorithms in identifying the dynamical states. 

Here, we describe the skeleton of the entire investigation. Starting with a given set of initial conditions and other control parameters (depending on the employed models) for four different coupled dynamics on network models, we evolve the individual model using the prescribed evolution rule discussed in the Results section. After an initial transient, we generate the time series data for each coupled dynamics on network model. We then have taken spatial profile of the coupled units at a particular time (henceforth described as snapshots) and characterize the snapshots into three different regimes, i.e. chimera, coherent and incoherent by the help of the reported behaviour of the system in the literature as well as by visually inspecting the dynamical behaviour of all coupled units (or nodes). Such collection of snapshots were generated for each coupled dynamics on networks for different parameter values. To summarize, the collection or the datasets are consisted of snapshots for all four coupled dynamics on network models collected at different time-points as well as different control parameters (depending on the underlying model) categorised in three distinct dynamical behaviours described above. For different models, the datasets was divided into two parts. The first part was used to train various machine learning algorithms for characterization of three different regimes. Thereafter, the another part of the dataset is used for testing purpose to check the accuracy of the trained algorithms.

The paper is organized as follows. After the first introductory section containing motivation, research gap, proposed plan and skeleton of the framework, in the second section we provide detailed discussions on various machine learning techniques employed by us. The third section contains the results as well as discusses their implications. The fourth section concludes the work followed by a section discussing future directions.


\section{Machine Learning Techniques}
In this section, we elaborate the five machine learning algorithms (RaF, MPRaF-T, MPRaF-P, MPRaF-N and RVFL-AE) that have been used here for identification and characterization of the spatial patterns.

\subsection{Random Forest}
Originally given by Breiman~\cite{breiman2001random}, random forest (RaF) is an ensemble of decision trees based on the concept of  bagging and random subspaces. Each decision tree of the RaF is trained on the randomly initialized vectors sampled independently with the same distribution of the original training dataset. Combining the concepts of bagging and random subspaces leads to an increase of the variance among the base classifiers.  Each decision tree generated on the bootstrapped version of the training data evaluates the $``mtry"$  number of random subspace features to chose the best split based on the impurity gain. The $``mtry"$  parameter determines the number of split tests at each non-leaf node of the tree. 
Random perturbation of the training data via boosting and random selection of the node splits via $``mtry"$ parameter results in improved generalization performance.   
The feature which optimizes the impurity gain is chosen as the best split. 
 The details of RaF are given in Algorithm \ref{tab:RaF}.  
\begin{algorithm}[H]
\caption{Random Forest Algorithm \cite{breiman2001random}}
    \label{tab:RaF}
\begin{algorithmic}
   \State{Training Phase:}\\
    Given: 
     \State{$X=M \times n$ is the training data with $M$ number of samples each of dimension $n$.}
    \State{$Y=M \times 1$ are labels of the training data.}
    \State{$R$ represents the ensemble size i.e. number of trees.}
    \State{Each tree in the random forest is represented as $T_{i}$, where $i=1, \dots ,R$.}
    \State{$`` mtry"$ is the number of randomly selected subset of features.}
    \State{$``minleaf"$ is the maximum number of samples in an impure node.}\\
    \begin{enumerate}
        \item Each tree $T_{i}$ is build using the bootstrapped versions of the training data $X$ with replacement.
        \item At each non-leaf node, the best feature split is selected among the $``mtry"$ randomly selected features from the training data.
        \item Repeatedly execute step 2 until one of the conditions is met:
        \begin{itemize}
            \item Node becomes pure.
            \item Node contains number of samples less than or equal to $minleaf$.
        \end{itemize}
    \end{enumerate}
    \vspace{1mm}
    Classification Phase:\\
    For each testing sample, each tree in the forest assigns the vote to a given sample data. Then based on the maximum voting, each data sample is assigned a class.
\end{algorithmic}
\end{algorithm}

\subsection{Oblique Random Forest via MPSVM}
RaF generates the decision trees based on the evaluation of features at each node of the tree and the final split occurs with the single feature that improves the purity of the node.
 However, this single feature based decision trees may not be optimal as they fail to capture the geometric structure of the data. To overcome this limitation, oblique decision tree ensembles were proposed. Unlike axis-parallel trees, oblique decision trees uses multiple features for each split. In 
 oblique decision tree ensemble via MPSVM \cite{zhang2014oblique}, splitting at each node is  based on the hyperplanes generated via multisurface proximal support vector machines (MPSVM). These hyperplanes capture the geometric structure of the data and hence show better performance as compared to the RaF. However, MPSVM is originally designed for binary class problems. Hence, Algorithm (\ref{tab:multi2binary}) is used to handle the multiclass problems. In this Algorithm $2$, Bhattacharyya distance is used to obtain the groups of maximum distance. The Bhattacharyya distance between the normal classes  $L_j$ and $L_k$,
$N(\eta_j,\sum_j)$ and $N(\eta_k,\sum_k)$, is given as 

\begin{equation}
    \label{eqn:Bhattacharyya distance}
        B(L_j,L_k)=\frac{1}{8}(\eta_k-\eta_j)^T\Bigg( \frac{\sum_j+\sum_k}{2}\Bigg)^{-1}(\eta_k-\eta_j)+ \frac{1}{2}ln\frac{|(\sum_j+\sum_k)/2|}{\sqrt{|\sum_j||\sum_k|}},
\end{equation}
 where $\eta_j, \eta_k$ and $\sum_j, \sum_k$  are the means and covariances of the distributions, respectively.
\begin{algorithm}[H]
 \caption{Multiclass to Binary class \cite{zhang2014oblique}}
    \label{tab:multi2binary}
\begin{flushleft}
\textbf{Input:}\\$X=M \times n$ is the training data with $M$ number of samples each of dimension $n$.\\
    $Y=M \times 1$ are labels of the training data.\\
       $\{L_1,\dots,L_c\}$ is the set of data labels.\\
    \textbf{Output:} $G_p$ and $G_n$ are the two hyperclasses or groups.
\end{flushleft}
    \begin{algorithmic}[1]
    
    \For{
        $j=1,\dots,c$}
        \State{Calculate the Bhattacharyya distance between every pair of classes $L_j$ and $L_k$, with $~~~~~~~~~~~k=j+1,\dots,c$ using equation (\ref{eqn:Bhattacharyya distance}).}
              \EndFor 
        \State{Chose classes $L_p$ and $L_n$ such that $L_p$ and $L_n$ are at largest Bhattacharyya distance, put them in groups $G_p$ and $G_n$, respectively.}
        \State{For every other class, if $B(L_k,L_p)<B(L_k,L_n)$ then assign the $L_k$ to the group $G_p$ else put in the group $G_n$.}
    \end{algorithmic}
\end{algorithm}

 Details of the MPRaF are given in the Algorithm \ref{tab:MPRaF}.  
\begin{algorithm}[H]
\caption{Oblique decision tree ensemble via MPSVM  \cite{zhang2014oblique}}
    \label{tab:MPRaF}
\begin{algorithmic}
   \State{Training Phase:}\\
    Given: 
     \State{$X=M \times n$ is the training data with $M$ number of samples each of dimension $n$.}
    \State{$Y=M \times 1$ are labels of the training data.}
    \State{$K$ represents the number of classes in the training data.}
    \State{$R$ represents the ensemble size i.e. number of trees.}
    \State{Each tree in the random forest is represented as $T_{i}$, where $i=1, \dots ,R$.}
    \State{$`` mtry"$ is the number of randomly selected subset of features.}
    \State{$``minleaf"$ is the maximum number of samples in an impure node.}\\
    \begin{enumerate}
        \item[1.] Each tree $T_{i}$ is build using the bootstrapped versions of the training data $X$ with replacement.
        \item[2. ] \textbf{If} $K>2$  at a node \textbf{then}
         \item[ ]      Group the data based on the Bhattacharyya distance into two maximal separated groups ($G_p, G_n$) using Algorithm \ref{tab:multi2binary}.
        \item[3.] At each non-leaf node, train  MPSVM based on the $``mtry"$ randomly selected features from the training data samples of the grouped data.
        \item[4.] Repeatedly execute Step 2 and Step 3 until one of the conditions is met:
        \begin{itemize}
            \item Node becomes pure.
            \item Node contains number of samples less than or equal to $minleaf$.
        \end{itemize}
    \end{enumerate}
    \vspace{1mm}
    Classification Phase:\\
    For each testing sample, each tree in the forest assigns the vote to the given sample data. Then based on the maximum voting, each data sample is assigned a class.
\end{algorithmic}
\end{algorithm}
While solving the generalized eigenvalue problem in MPSVM (Step-3 of the Algorithm-\ref{tab:MPRaF}), the matrices appearing in the formulation of MPSVM may be ill conditioned, i.e., they may be positive semi-definite. To handle this singularity problem in the MPSVM formulation, different regularization techniques are used: $1)$ Tikhonov regularization \cite{R49} $2)$ Axis-parallel split regularization \cite{mangasarian1979nonlinear,evgeniou2000regularization} and $3)$ Null Space regularization \cite{chen2000new}. In the Tikhonov regularization method, a small constant is added to the diagonal entries of the matrix to be regularized. Suppose $G$ is rank deficient, then $G$ is regularized as
\begin{align}
    G'=G+\delta \times I,
\end{align}
while as in axis-parallel split regularization, if a matrix becomes singular, then the axis parallel split is used to continue the generation of a tree.
In the Null space regularization method, orthogonal projections are used for regularization. 
Based on the different approaches used to handle the singularity problem, the classification algorithms are named accordingly.  
MPRaF-T, MPRaF-P, and MPRaF-N represent the
MPSVM-based RaFs with Tikhonov, axis-parallel, and NULL
space regularization, respectively. In the study \cite{zhang2014oblique}, different values for the \textit{``minleaf"} were evaluated and concluded that lower the \textit{``minleaf" }better the average rank, hence \textit{`` minleaf"}$=1$ is the reasonable choice.

\subsection{Sparse Pre-Trained Random Vector Functional Link (SP-RVFL/ RVFL-AE) network}
Consider the training dataset $(X,Y)$ where $X$ represents the input features and $Y$ are the corresponding labels of the data samples. Let each sample $x_i\in \mathbb{R}^d$  and $N$ represents the number of hidden neurons, then for each output node there are $d+N$ input connections. Unlike standard RVFL network \cite{pao1994learning} which generates the hidden layer weights randomly, SP-RVFL \cite{zhang2019unsupervised} learns the hidden layer weights via $l_1$-regularized autoencoder. The objective function of the $l_1$-regularised autoencoder is given as
\begin{align}
\label{eqn:7}
    \underset{\hat{w}}{min} ~~\norm{\hat{H}\hat{w}-X}^2+\norm{\hat{w}}_1,
\end{align}
where $X, \hat{H}$ and $\hat{w}$ represent the input features, hidden layer weights generated randomly and learned  output weights by the autoencoder, respectively. The objective function (\ref{eqn:7}) is solved via fast iterative shrinkage thresholding algorithm (FISTA) \cite{FISTA}. The weights $(\hat{w})$ learned via optimizing the objective function (\ref{eqn:7}) are used to initialize the hidden layer weights of the standard RVFL network \cite{pao1994learning}. The hidden layer biases are chosen as
\begin{align}
    \hat{b_i}=\frac{\sum_{j=1}^d\hat{w}_{ij}}{d}, ~~i=1,2,...,N. 
\end{align}

With the learned parameters $(\hat{w}, \hat{b})$, the hidden layer output of RVFL-AE is given as follows:
\begin{align}
    H=g(X\hat{w}+\hat{b}),
\end{align}
where non-linear activation function is given by $g(.)$.
Based on the original and hidden layer output, the extended feature matrix $M=[H,X]$ is constructed. The objective function  optimized by RVFL-AE is given as:
\begin{align}
\label{eqn:stdRVFL}
    \underset{\theta_s}{min}~~ \norm{M\theta_s-Y}^2+\lambda \norm{\theta_s}^2,
\end{align}
where $\theta_s$ are the output layer weights, $Y$ are target labels, and $\lambda$ is the regularisation parameter. The objective function $(\ref{eqn:stdRVFL})$ can be solved either by Moore-Penrose pseudoinverse ($\lambda=0$) \cite{} or via regularized least squares (ridge regression) with $\lambda\neq 0$. The solution via Moore-Penrose pseudoinverse  is given by: $\theta_s=M^+Y$, where $M^+$ denotes the pseudoinverse of $M$, while as with ridge regression the solution is given as:
\begin{align}
    \text{Primal Space}:~~\theta_s=(M^TM+\lambda I)^{-1}M^TY ,\\
    \text{Dual Space:}~~ \theta_s=M^T(MM^T+\lambda I)^{-1}Y.
\end{align}

For more details, interested readers are referred to \cite{zhang2019unsupervised}.

\section{Results and Discussions}
This article aims at characterizing different dynamical states using the machine learning algorithms described in the previous section. We divide the validation of the algorithms into four parts by considering two time-discrete and two time-continuous dynamical models to generate the required data for the training and the testing. It should be noted that the datasets used for the training and the testing the dynamical models are primarily the snapshot profiles i.e., a $[N \times 1]$ vector, representing the state of the chosen model at a particular time-step in the steady state (i.e. after initial transient) with $N$ being the size of the network. A point to note here is that the sizes of the different networks chosen for different models stem from the corresponding literature from which all the control parameter is adapted to generate snapshots with known dynamical behavior. We also visually inspect each of the snapshots to confirm its label as a coherent, incoherent, or chimera state. Although, in few cases, the frequency or the amplitude is used to discern the chimera and other states, the use of snapshot profile enables us to perform a generalized approach regardless of the time-continuous or time discrete nature of the considered system. We argue that because of the approach undertaken in the manuscript, this  technique can be universally applicable to any coupled dynamics model and real-world time series data having chimera patterns. 

We hereafter, first briefly describe spatial behaviour of the chimera and other states followed by a demonstration of occurrence of these states for different dynamical models. Chimera state is described as a hybrid state where the coherent and the incoherent dynamics co-exist together in a network of identical oscillators. The interaction pattern of the nodes in the network is encoded by an adjacency matrix $\mathcal{C}$ such that $c_{ij} = 1$, if a link exists between a pair of the nodes $i$ and $j$, and zero, otherwise. Here, throughout the article, we have considered an un-directed (i.e. adjacency matrix $\mathcal{C}$ is symmetric) single-layer regular network ($\mathcal{C}(N,r)$; $S^1$ : ring network), where all the $N$ nodes have the same node degree $k$ which is also represented as coupling radius $r$, defined as $r=k/2N$. 

Formally, we define the existence of coherence in the snapshot profile of coupled dynamical model on a network as~\cite{omelchenko2011loss} 
\begin{equation}
  \lim\limits_{N \rightarrow \infty} \lim\limits_{t \rightarrow \infty}\sup\limits_{i,j  \in U_{\xi}^N (x)} \mid{z_i(t)-z_j(t)} \mid \rightarrow 0 \, \, \text{for} \, \, \xi\; \rightarrow 0,
  \label{eq.cohr}
  \end{equation}
  where $U_{\xi}^{N} (x) = \{ j : 0 \le j \le N, \mid{\frac{j}{N} - x} \mid <  \xi \}$ represents the network neighborhood from any point $x \in S^1$ ( i.e. regular network), spanning spatial distance ($\xi$). The $ \mathbf{Z}(t) \in \mathbb{R}^N$ represents the state vector $(N\times1)$ with the components $z_i(t) \ni \mathbf{Z}=\{ z_1,z_2,\dots,z_N\}$, as a real dynamical variable at time $t$ for the $i^{th}$ node for the considered model. $N$ stands for the number of nodes in the network and hence the adjacency matrix is of dimension $N \times N$ and the state vector is of dimension $N \times 1$. Henceforth, we denote the state vector $\mathbf{Z}(t)$ for different models as dynamical state of the underlying system. The snapshot profile ($ \mathbf{Z}(t)$) considered as input for ML algorithms  for each model system is the dynamical state of the model captured at a specific time. In the asymptotic limit ($N \to \infty$), $\mathbf{Z}(t)$ approaches a smooth spatial profile, without any discontinuity, for a coherent state. Furthermore, in the extreme case (i.e. for high coupling range), all components of $\mathbf{Z}(t)$ assume the same value and thus producing a complete coherent state denoted by a flat spatial profile. On the other hand, the spatial profile of an incoherent state does not show any smooth segment in the snapshot. A chimera state, which comprises of coexisting coherent and incoherent states, shows smooth profiles broken by non-smooth segments representing discontinuities in the snapshot. A detailed discussions on types of the chimera state can be found in Ref.~\cite{ghosh2019taming}. In the following, we describe the governing equation for the time evolution of the state vector for different dynamical models and for regular single-layer un-directed network.

\begin{figure}[t]
\centerline{\includegraphics[width=\columnwidth]{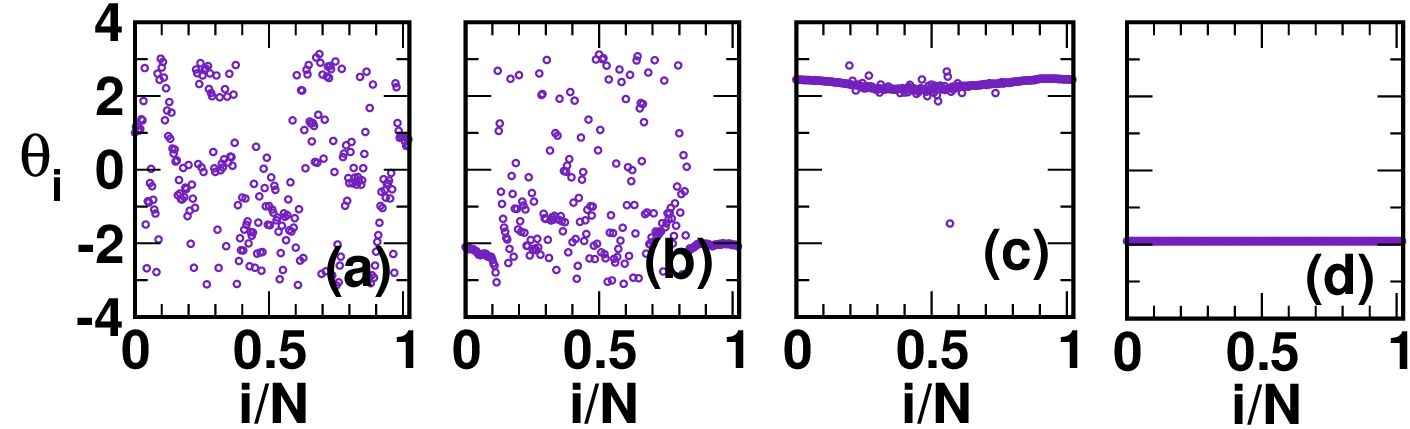}}
\caption{Snapshots of phase-state of non-locally coupled Kuramoto oscillators arranged in a regular network $\mathcal{C}(N,r)$ demonstrating (a) Incoherent state for $\alpha=1.75$, (b) Chimera state for $\alpha=1.51$, (c) Chimera state for $\alpha=1.39$, and (d) Coherent state $\alpha=0.9$. Other parameters are $N=256$, $r=0.32$, $\lambda=0.1$, $\omega_i=\omega=0.01, \forall_i$.}
\label{fig1_kosc}
\end{figure}

\subsection{Coupled Kuramoto Oscillators}
The famous coupled Kuramoto model~\cite{kuramoto1975kuramoto,strogatz2000kuramoto} is one of the most explored coupled dynamical system which despite of being very simple in nature, is capable of exhibiting several complex dynamical phenomena. A plethora of work has been done demonstrating its applicability in diverse areas of research~\cite{rodrigues2016kuramoto} including the first report of emergence of chimera~\cite{kuramoto2002coexistence,abrams2004chimera}.

The state vector for the coupled Kuramoto oscillators model is defined in terms of the phases ($\theta$) of $N$ nodes of the network. Therefore, $\Theta={\theta_1,\theta_2,\dots,\theta_N}$ represents the dynamical state of the system at a particular time, representing the phase of each node as a real variable $\theta_i\in \mathbb{R}, \forall i=1,...,N$. The governing equation for the Kuramoto model utilizing the adjacency matrix can be written as~\cite{sarika2017repulsion} 
\begin{equation}
    \dot{\theta_i} = \omega_i - \frac{\lambda}{\sum_{j} c_{ij}} \sum_{j=1}^{N} c_{ij} \sin(\theta_j - \theta_i + \alpha),    
    \label{KOSC}
\end{equation} 
where $\lambda$ denotes the coupling strength, $\omega_i =\omega; \forall_i$ denotes the constant natural frequency of all the oscillators and $\alpha$ denotes a constant phase lag. We choose the system parameters, namely, $\alpha$ and $\lambda$ as well as use a specially constructed hump-back initial function (as described in Ref.~\cite{abrams2004chimera,Abrams2006a}) such that varying the value of  the phase lag parameter $\alpha$ gives rise to different dynamical states. Fig.~\ref{fig1_kosc} depicts different dynamical states arising from the choice of the phase lag parameter $\alpha$. Note that Figs.~\ref{fig1_kosc} (b) and (c) demonstrating chimera states comprise of smooth segments broken by incoherent regions, whereas Fig.~\ref{fig1_kosc} (a) does not have a smooth segment denoting an incoherent profile and Fig.~\ref{fig1_kosc} (d) shows a completely smooth profile representing a coherent state.
\begin{figure}[t]
  \centerline{\includegraphics[width=\columnwidth]{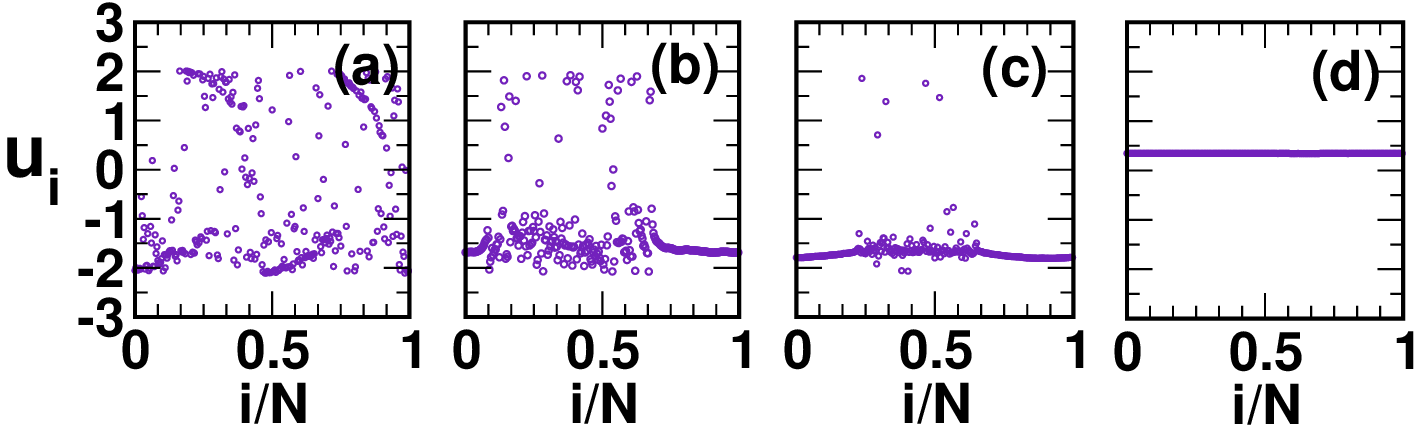}}
  \caption{Snapshot profiles of activator variable of  non-locally coupled FHN oscillators arranged in a regular network $\mathcal{C}(N,r)$ demonstrating (a) Incoherent state for $\phi=\pi/2 +0.35$, (b) Chimera state for $\phi=\pi/2 - 0.1$, (c) Chimera state $\phi=\pi/2 - 0.15$, and  (d) Coherent state $\phi=\pi/2 - 0.35$. Other parameters are $N=300$,$r=0.35$,$\lambda=0.1$,$a=0.5$,$\varepsilon=0.05$. }
  \label{fig2_fhn}
  \end{figure}

\subsection{Coupled FitzHugh-Nagumo (FHN) Oscillators}
Next, we consider FitzHugh–Nagumo (FHN) model which has been a famous model for neural excitability~\cite{masoliver2017fhn}. This two-dimensional coupled dynamical system demonstrates chimera states for networks consisting of coupled oscillatory~\cite{omelchenko2013nonlocal} and excitatory~\cite{Semenova2016fhn} FHN nodes. Note that, henceforth, we use ``FHN" as abbreviation of name ``FitzHugh–Nagumo" throughout the manuscript. The governing equation for this model can be described as~\cite{schulen2019sol} 
\begin{equation}\
\label{FHN}
\begin{array}{c}
\varepsilon\frac{du_i}{dt}=u_i-\frac{u^3_i}{3}-v_i 
+\frac{\lambda}{\sum_j c_{ij}}\sum\limits_{j=1}^{N} c_{ij} \{b_{uu} (u_j -u_i) 
+ b_{uv} (v_j - v_i)\} \; ,\\
\frac{dv_i}{dt}=u_i + a + \frac{\lambda}{\sum_j c_{ij}}\sum\limits_{j=1}^{N} c_{ij}
\{b_{vu} (u_j - u_i) 
+ b_{vv} (v_j - v_i)\} \; .
\end{array}
\end{equation}
The state vector of this model can be described by two variables namely $u_i$ and $v_i$ representing the activator and inhibitor variables, respectively. A small parameter responsible for the time scale separation of the fast activator and the slow inhibitor is given by $\varepsilon > 0$. Here we fix $\varepsilon=0.05$. System parameter $a$ defines the excitability threshold. For a single FHN unit it decides between the excitable state for $(|a| > 1)$, or the oscillatory state for $(|a| < 1)$. Here, we consider the oscillatory regime and fix $a=0.5$. The initial state is randomly chosen from a circle, $u^2 + v^2 =4$. Furthermore, the FHN model, considered here, includes direct as well as cross couplings between activator $u_i$ and inhibitor $v_i$ variables, which is encoded by a rotational coupling matrix~\cite{omelchenko2013nonlocal,schulen2019sol} 
\begin{equation}
B = \begin{pmatrix}
b_{uu} & b_{uv} \\
b_{vu} & b_{vv}
\end{pmatrix}
=
\begin{pmatrix}
cos \phi & sin \phi \\
-sin \phi & cos \phi
\end{pmatrix}.
\end{equation}
Different dynamical states for the model can be demonstrated by choosing different rotation angle $\phi$ for the coupling matrix as depicted in Fig.~\ref{fig2_fhn}. Further details about various choices of the parameters can be explored in Ref.~\cite{schulen2019sol} and references therein. The Figs.~\ref{fig2_fhn} (a) and (d) present a coherent and incoherent state, respectively for different values of the rotating angle $\phi$ keeping other system parameters the same. The Chimera state demonstrated in the Figs.~\ref{fig2_fhn} (b) and (c)  for such systems has been previously reported in Ref.~\cite{omelchenko2013nonlocal}. 
\begin{figure}[t]
  \centerline{\includegraphics[width=\columnwidth]{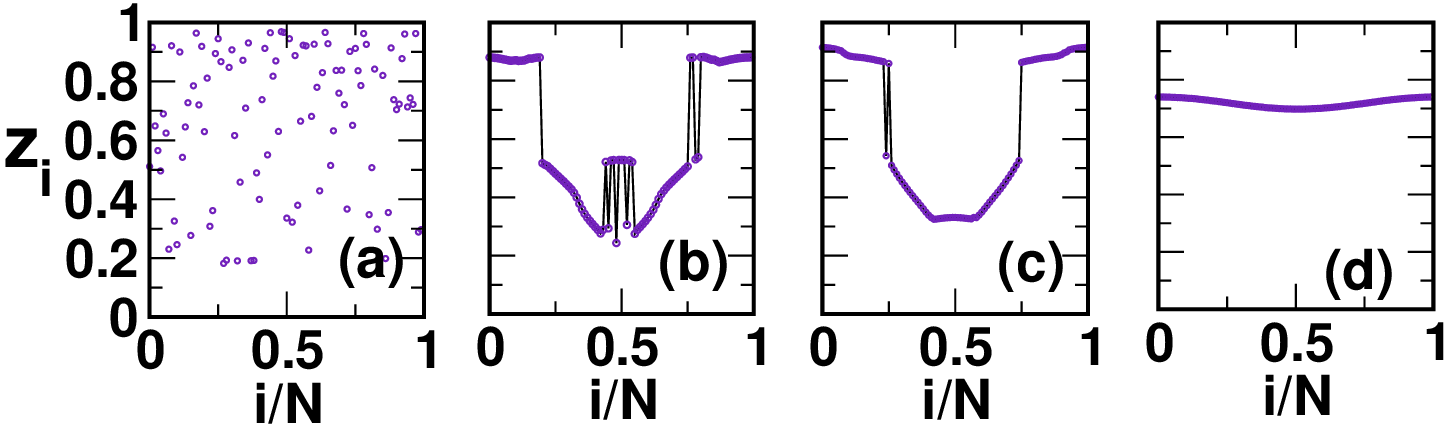}}
  \caption{Snapshot profiles of state variable of non-locally coupled logistic map arranged in a regular network $\mathcal{C}(N,r)$ demonstrating (a) Incoherent for $\varepsilon=0.1$, (b) Chimera state for $\varepsilon=0.34$, (c) Chimera state for $\varepsilon=0.37$, and (d) Coherent state for $\varepsilon=0.7$. Other parameters are  $N=100$, $r=0.32$, $\mu=4.0$.}
  \label{fig3_logs}
  \end{figure}

\subsection{Coupled Logistic Map}
Next, we consider the logistic map model represented by $f(z)=\psi z (1-z)$ where the state variable ${z}\in[0,1]$ as local dynamics for each node ~\cite{may1976logs}. We have considered the logistic map in its chaotic regime and therefore set the bifurcation parameter $\psi=4.0$~\cite{ott1993logs}. This simple model has been thoroughly explored  by the non-linear dynamics community to understand diverse spatio-temporal phenomena in a wide range of real-world complex systems~\cite{logs_appl} among which chimera has also been shown in both the single layer~\cite{omelchenko2011loss} and multiplex networks~\cite{ghosh2016emergence}. The dynamical evolution of the coupled logistic maps model is governed by
\begin{equation}
z_i^{t+1}=f(z^t_i)+\frac{\varepsilon}{\sum_{j} c_{ij}} \sum_{j=1}^{N} c_{ij}[ f(z^t_j) - f(z^t_i) ] ,
\label{log.map}
\end{equation}
where $\varepsilon \in [0,1]$ depicts the overall coupling strength and $\sum_{j} c_{ij}$ is the normalizing factor. Depending on the coupling strength $\varepsilon$ value and a special initial condition (Ref.~\cite{ghosh2016emergence}), different dynamical states can arise for the coupled logistic map model as shown in Fig.\ref{fig3_logs}.  Figs.\ref{fig3_logs} (b) and (c) report chimera state for mid-range coupling strength~\cite{omelchenko2011loss,ghosh2016emergence} whereas Figs.\ref{fig3_logs} (a) and (d) demonstrate the incoherent and the coherent state for low and high coupling strength, respectively.
\begin{figure}[t]
  \centerline{\includegraphics[width=\columnwidth]{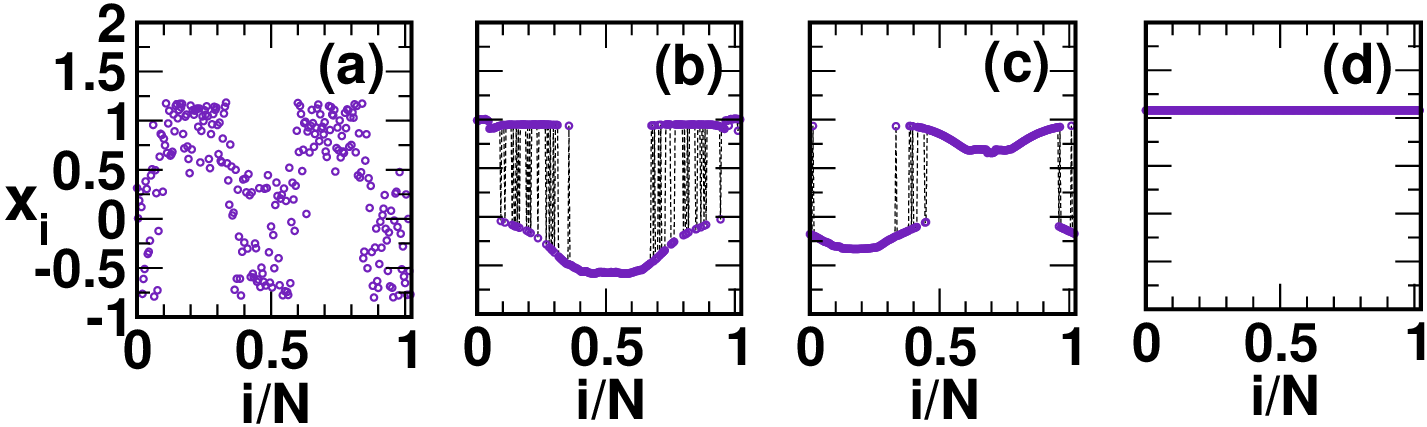}}
  \caption{Snapshot profiles of $x$ state variable of non-locally coupled H\'enon maps on regular network $\mathcal{C}(N,r)$ demonstrating (a) Incoherent state for $\varepsilon=0.1$, (b) Chimera state for $\varepsilon=0.26$ (c) Chimera state for $\varepsilon=0.3$, and (d) Coherent state for $\varepsilon=0.7$. Other parameters are $N=256$,  $r=0.32$, $\alpha=1.4$, $\beta=0.3$.}
  \label{fig4_henn}
  \end{figure}
\subsection{Coupled H\'enon Map}
Finally, we consider another well-known time-discrete model known as H\'enon map~\cite{henon1976} represented as $\mathcal{H}(x,y) \equiv (1 - \alpha x^2 + y; \beta x$) for dynamical evolution for each node. Here also, we set the system parameters $\alpha=1.4$ and $\beta=0.3$ to obtain chaotic local dynamics. The coupled H\'enon maps is a well-studied model which is known to demonstrate different complex behaviors including emergence of hybrid spatio-temporal patterns of chimera state~\cite{semenova2015does}. The time evolution of the map along with the adjacency matrix can be described as~\cite{ghosh2019taming}
\begin{equation} \label{hp_evoleq}
	\begin{split}
		x_i^{t+1} &=f(x_i^t,y_i^t) +\frac{\varepsilon}{\sum_{j} c_{ij}} \sum_{j=1}^{N} c_{ij}[ f(x_j^t,y_j^t)-f(x_i^t,y_i^t) ], \\
		y_i^{t+1} &= \beta x_i^t,
	\end{split}
	\end{equation}
where $f(x_i^t,y_i^t)= 1 - \alpha (x_i^t)^2 + y_i^t$, $\varepsilon \in [0,1]$ presents the overall coupling strength and $\sum_{j} c_{ij}$ is the normalizing factor. The initial condition is chosen as described in Ref.~\cite{semenova2015does,ghosh2019taming}. Similarly as in the case of logistic map, the choice of the value of coupling strength $\varepsilon$ causes different dynamical states to arise in coupled h\'enon map model as depicted in Fig.\ref{fig4_henn}.  Figs.\ref{fig4_henn} (b) and (c) show chimera state for mid-range coupling strength~\cite{semenova2015does} and Figs.\ref{fig4_henn} (a) and (d) demonstrate the incoherent and the coherent state for low and high coupling strength, respectively.

\subsection{Setup of ML parameters}
The different parameters corresponding to different algorithms were set as follows:
 the number of trees in each forest is $100$ for RaF, MPRaF-T, MPRaF-P and MPRaF-N,
 the regularization parameter in MPSVM, $\delta=0.01$, $minleaf=1$, $mtry=round(\sqrt{n})$, where $n$ is the number of features of a sample
  , the number of hidden neurons for RVFL-AE was chosen from the range $2:20:302$ and radbas activation function was used in RVFL-AE. 

\begin{table*}
  \centering
  \adjustbox{width=\textwidth}{
  \begin{tabular}{|l|c|c|c|c|c|}
  \hline
  \textbf{Model}&\textbf{RVFL-AE}&\textbf{RaF}&\textbf{MPRaF-T}&\textbf{MPRaF-P}&\textbf{MPRaF-N}\\
  \hline
Model-Kuramoto&0.997&0.9487&1&0.9518&0.9996\\
Model-Logistic-Map&1&0.9813&0.9773&0.972&0.7973\\
Model-Hennon-Map&1&0.9733&0.9933&0.9773&0.9947\\
Model-FHN&1&0.986&0.9973&0.9933&1\\
\hline
  \end{tabular}}
  \caption{Prediction accuracy (\%) for the training data corresponding to different models by different classification algorithms.}
  \label{tab:train_acc}
\end{table*}
\begin{table*}
   \centering
   \resizebox{0.70\textwidth}{!}{
   \begin{tabular}{|l|c|c|c|c|c|c|c|}
  \hline
  \multicolumn{8}{|c|}{Model-Kuramoto}\\
  \hline
  \multicolumn{8}{|c|}{Train-Size$=900 \times 256$, Test-Size$= 100\times256$}\\
  \hline
Discarded=0&&\multicolumn{3}{c}{Sensitivity}&\multicolumn{3}{|c|}{Specificity}\\
 \hline
Algorithm&Accuracy&Chimera&Coherent &InCoherent&Chimera&Coherent &InCoherent\\
\hline
MPRaF-N&1&1&1&1&1&1&1\\
MPRaF-P&0.97&0.925&1&1&1&0.9571&1\\
MPRaF-T&1&1&1&1&1&1&1\\
RaF&0.96&0.9&1&1&1&0.9429&1\\
RVFL-AE&0.93&0.825&1&1&1&0.9&1\\
\hline
\multicolumn{8}{|c|}{Model-Logistic-Map}\\
\hline
\multicolumn{8}{|c|}{Train-Size$=300 \times 100$, Test-Size$= 101\times 100$}\\
\hline
Discarded=8&&\multicolumn{3}{c}{Sensitivity}&\multicolumn{3}{|c|}{Specificity}\\
\hline
Algorithm&Accuracy&Chimera&Coherent &InCoherent&Chimera&Coherent &InCoherent\\
\hline
MPRaF-N&0.8817&0.7083&1&0.8095&0.971&0.8&1\\
MPRaF-P&0.8925&0.9583&0.8125&1&0.942&0.9778&0.9306\\
MPRaF-T&0.9032&0.875&0.8958&0.9524&0.913&0.9778&0.9722\\
RaF&0.914&0.875&0.9167&0.9524&0.9275&1&0.9583\\
RVFL-AE&0.7312&0.7917&0.7917&0.5238&0.942&0.7556&0.8611\\
\hline
\multicolumn{8}{|c|}{Model-Hennon-Map}\\
\hline
\multicolumn{8}{|c|}{Train-Size$=300 \times 256$, Test-Size$= 101\times256$}\\
\hline
Discarded=5&&\multicolumn{3}{c}{Sensitivity}&\multicolumn{3}{|c|}{Specificity}\\
\hline
Algorithm&Accuracy&Chimera&Coherent &InCoherent&Chimera&Coherent &InCoherent\\
\hline
MPRaF-N&0.8333&0.4545&1&0.8261&0.9459&0.7333&1\\
MPRaF-P&0.8229&0.8636&0.8039&0.8261&0.8919&0.9333&0.9178\\
MPRaF-T&0.7813&0.2273&1&0.8261&0.9595&0.6222&0.9863\\
RaF&0.8021&0.9091&0.7647&0.7826&0.8378&0.9556&0.9315\\
RVFL-AE&0.5938&0.2727&0.8824&0.2609&0.8243&0.4222&1\\
\hline
\multicolumn{8}{|c|}{Model-FHN}\\
\hline
\multicolumn{8}{|c|}{Train-Size$=300 \times 300$, Test-Size$= 100\times300$}\\
\hline
Discarded=3&&\multicolumn{3}{c}{Sensitivity}&\multicolumn{3}{|c|}{Specificity}\\
\hline
Algorithm&Accuracy&Chimera&Coherent &InCoherent&Chimera&Coherent &InCoherent\\
\hline
MPRaF-N&0.8763&0.7447&1&1&1&0.9167&0.9167\\
MPRaF-P&0.6701&0.3191&1&1&1&0.5556&1\\
MPRaF-T&0.7113&0.4043&1&1&1&0.6389&0.9722\\
RaF&0.6804&0.3404&1&1&1&0.5833&0.9861\\
RVFL-AE&0.7526&0.4894&1&1&1&0.6667&1\\
\hline
  \end{tabular}}
  \caption{Prediction accuracy (\%), sensitivity and specificity  for the test data corresponding to different models by different classification algorithms.}
  \label{tab:test_acc}
\end{table*}

\subsection{Implementation of ML techniques}
We show that the ML algorithm identifies the chimera patterns for all the dynamical models and thus establishes the universality of the proposed technique. We generate two data sets namely (training and testing) for each model described in the previous subsections for our investigation. The first data set is the ``labeled" training data where each snapshot is described as either coherent, chimera or incoherent in nature. This data set is generated from the range of system's parameter reported in literature to show a particular dynamics (Coherent, Chimera or Incoherent). We have performed visual inspection of the snapshot profiles and a set of them is represented in the Figures~\ref{fig1_kosc},\ref{fig2_fhn},\ref{fig3_logs},\ref{fig4_henn}, corresponding to different dynamics for different models (with the corresponding systems parameter mentioned in the captions). Note that, the labels at this stage is assigned based on the parameters which are well known in the literature to produce a particular type of dynamical state. We have trained our ML algorithms using this data set. Next, we prepared the testing data which is generated from the snapshot profiles by randomly choosing the system's parameters. We used all the trained ML algorithms to predict the nature of the profiles for the testing data set. We then visually inspected the states to prepare the ``ground truth labels" for the states and compared them with the labels predicted by the ML algorithms. The results and accuracy of the comparisons are presented in the next subsection demonstrating the efficiency of the ML approach. The size of training and testing data samples is given in Table (\ref{tab:test_acc}). Here, the training and testing data samples are represented as \textit{Number of samples} $\times$ \textit{Number of features}. Note that the mentioned test size includes the discarded states~\cite{disputed} corresponding to each model.

Note that for classification problems in the machine learning field, one uses two data sets, one for the training of algorithm and another for evaluation of the algorithm, respectively. The training data is used by the algorithm to generate the hypothesis for the classification, i.e., using the training data, a algorithm learns the data distribution. The hypothesis generated by the algorithm based on the training data is then evaluated on the test data. How good the algorithm performs on the test data (also known as unseen data) determines the efficiency of the algorithm. Thus, we are not concerned about the performance of the algorithm on the training data and focus only on the testing data. 
According to the {\it No-Free-Lunch theorem} \cite{no_free_lunch} the best classier will not be the same for all the data sets. Hence, the performance of the different algorithms may vary across different data sets.

\subsection{Prediction Accuracy corresponding to different dynamical states}
The measures used to evaluate the performance of the algorithms are sensitivity, specificity and accuracy. 
\begin{table}[]
    \centering
    \caption{Confusion matrix for multi-class classification problem.}
    \label{tab:Confusion_matrix}
   \begin{tabular}{l|l|c|c|c|}
\multicolumn{2}{c}{}&\multicolumn{3}{c}{True Class}\\
\cline{3-5}
\multicolumn{2}{c|}{}&A&B&C\\
\cline{2-5}
\multirow{2}{*}{Predicted Class}& A & $TP_A$ & $E_{BA}$ &$E_{CA}$\\
\cline{2-5}
& B & $E_{AB}$ & $TP_B$ &$E_{CB}$\\
\cline{2-5}
& C & $E_{AC}$ & $E_{BC}$ &$TP_C$ \\
\cline{2-5}
\end{tabular}
\end{table}
Table-(\ref{tab:Confusion_matrix}) represents the confusion matrix for a three class problem (Here, A,B and C represents the three classes). The diagonal entries are the correctly classified samples where as off-diagonal elements represent the miss-classifications. One can see from the Table-(\ref{tab:Confusion_matrix}) that $TP_A$ represents the number of $A$ class samples classified/predicted  as $A$-class by the classifier i.e. the number of correctly classified samples of class-$A$. $E_{BA}$ represents the number of $B$-class samples incorrectly classified as $A$-class and $E_{CA}$ is the number of $C$-class samples classified as $A$-class by the classifiers.
Thus, false negative in the $A$-class, $FN_A=E_{AB}+E_{AC}$ i.e. total number of $A$-class samples that were incorrectly classified as $B$ and $C$. Similarly, the false positive of class $A$, $FP_{A}=E_{BA}+E_{CA}$ i.e. total number of samples   that were incorrectly classified as $A$-class samples. Similarly, the true positive (TP), false positive (FP) and false negative (FN) of other classes can be defined. Now, in order to evaluate the performance of the algorithms, measures used are  accuracy, sensitivity and specificity. The details of each measure are given as:
\begin{align}
    \text{Accuracy}=\frac{TP_A+TP_B+TP_C}{TP_A+TP_B+TP_C+E_{BA}+E_{CA}+E_{AB}+E_{CB}+E_{AC}+E_{BC} }.
\end{align}

Sensitivity or True positive rate (TPR) of a classification algorithm is the ratio of the correctly classified positive samples of a class to the total number of negative samples of the class as given in equation (\ref{eqn:TPR}).
Specificity, True negative rate (TNR) is the ratio of correctly classified negative samples of a class to the total number of negative samples of a class as given in equation (\ref{eqn:FPR}).

\begin{align}
\label{eqn:TPR}
   \text{Sensitivity,}~ TPR_i=\frac{TP_i}{TP_i+FN_i} ,\\
    \label{eqn:FPR}
   \text{Specificity,}~ FPR_i=\frac{TN_i}{TN_i+FP_i},
\end{align}
where $i=A,B,C$ represents the concerned class. For more details about performance measures, interested readers are referred to the paper on classification assessment methods \cite{Classification assessment methods}.

We have trained the ML algorithms using the dynamical states as input and trained to classify the states into three categories namely coherent, incoherent and chimera states. 
The results presented in the Table (\ref{tab:train_acc}) represent the average training accuracy obtained via $5$-times four fold cross validation technique. One can see from the Table (\ref{tab:train_acc}), all algorithms
show more than $94\%$ accuracy, except in Model-Logistic-Map where MPRaF-N shows $79\%$ accuracy. 

\hspace{0.8cm}
Table (\ref{tab:test_acc}) represents the classification accuracy, sensitivity and specificity measures corresponding to the test data.
In Model-Kuramoto, all the algorithms achieved more than  $93\%$ accuracy. 
In Model-Logistic-Map, RVFL-AE shows lower performance as compared to the other given algorithms, however, RaF and MPRaF-T achieved more than $90\%$ accuracy while  MPRaF-N shows $88.17\%$ accuracy. 
In Model-H\'enon-Map, MPRaF-N and MPRaF-P achieved relatively better accuracy ($83.33\%$ and $82.29\%$ respectively)  as compared to the RaF and MPRaF-T algorithms which showed $80.21\%$ and $78.13\%$ accuracy, respectively. One can see that RVFL-AE shows lower performance ($59.38\%$ accuracy) as compared to the other baseline algorithms.
In Model FHN Osc, MPRaF-N shows relatively better performance ($87.63\%$ accuracy) followed by RVFL-AE  with $75.26\%$ and MPRaF-T with $71.13\%$ accuracy while RaF  and MPRaF-P achieved around $67\%$ accuracy. The sensitivity and specificity measures of different machine learning algorithms in different models corresponding to chimera, coherent, and incoherent classes are also given Table (\ref{tab:test_acc}).  

From the above analysis, one can conclude that the performance of the classification algorithms MPRaF-N, MPRaF-P, MPRaF-T and RaF are consistently better in 
the coupled Kuramoto oscillator model, the coupled Logistic maps model and the couppled Hennon maps model. Except in the coupled FHN oscillators model, where the performance of the RVFL-AE is relatively better, RVFL-AE shows relatively lower performance as compared to other algorithms. Note that, we have discarded the disputed states~\cite{disputed} while computing the success rate of the algorithms.

\section{Conclusion} 
To summarize, we have employed several machine leaning techniques to characterize different dynamical states arising due to the interaction between its constituent entities with a special focus on chimera states. We have used two time-discrete and two time-continuous models to validate our approach and found that the algorithms are quite successful in  characterizing the emergent states into chimera, coherent and incoherent states. A point to be noted here that all the selected models are well represented in literature to display chimera states. Here, we use them as a platform to demonstrate the utility of machine learning approach to advance the  automated characterization of the emergent dynamics. We have demonstrated that different algorithms show varying performance based on the dynamical states used. The identification of states in the coupled Kuramoto oscillators model by the classification models achieved more than $93\%$ accuracy. In
the coupled Logistic maps model, the classification algorithms (except RVFL-AE) achieved more than $88\%$ accuracy. 
In coupled Hennon maps model, RVFL-AE and MPRaF-T show relatively lower performance (around $59\%$ and $78\%$, respectively) while as other classification algorithms (MPRaF-N, MPRaF-P, RaF) achieved more than $80\%$ accuracy.
For all the coupled dynamics on networks models, except FHN, the  classification algorithms like RaF, MPRaF-N, MPRaF-P, and MPRaF-T are consistently performing better. However, for coupled FHN oscillators the performance of RaF and MPRaF-P is relatively lower as compared to their performance for other dynamical models. 
The appearance of the chimera state vary widely as reported in the vast range of chimera literature for different systems. In the present study, we have only considered time-static chimera states without adding a temporal aspect to our work for classification of chimera state whose appearance vary with time. For example, ND Tsigkri-DeSmedt {\it et al.}~\cite{DeSmedt2016,DeSmedt2018} reports a travelling chimera state in neuronal network with non-trivial profile or Suda {\it et al.}~\cite{Suda2018} reports a breathing multi-chimera states, which will be rather very difficult to categorise using the traditional approaches as well as our present methodology. Furthermore, the coherent state of a system itself can widely vary in different systems ranging from a sloppy smooth spatial profile to a completely flat spatial profile. While, we have included both the single/multi chimera states and the flat/sloppy coherent profiles in our sample set, a comparative analysis of the performance of the algorithms for intra-classification of a specific dynamics is absent. However, the goal of our present investigation is to provide a simple yet universal approach to identify chimera state and other dynamical profiles using the ML algorithms which pave the way for a complete new area of study for automated classification of various dynamical profile, as outlined in the next section, merging the field of coupled non-linear dynamics and machine learning.

\section{Future Perspectives}
This investigation demonstrates the use of machine learning techniques to characterize different dynamics and to identify the chimera state in a given time-discrete or time-continuous coupled dynamics on networks model. Chimera state has been found to play a crucial role in cognitive functions in human brains~\cite{bansal2019cognitive}. Electroencephalogram (EEG) readings at the onset of epileptic seizure is reported to show chimera-like patterns~\cite{andrzejak2016all}. However, various identification measures existing in literature, being designed in a model-specific way, lacks the universal applicability for the task as well as may require additional threshold criteria to correctly identify the chimera state. Our simplistic approach based on the machine learning techniques demonstrates high accuracy rate in identifying and characterizing such spatial patterns, providing new future ways to diagnose various neural disorders associated with such patterns. Moreover, our approach is not only limited to characterization of chimera patterns. Diverse emergent dynamics of various coupled complex systems exhibit a plethora of novel phenomena resulting in diverse applications in various fields ranging from power grid to social systems. For example, cluster synchronization, which represents a dynamic division of a network in two or more different groups of synchronized oscillators, commonly emerges in neural, social and many other real-world systems represented by networks~\cite{Palmigiano2012,Kanter2011,Blasius1999,Singh2017}. The ML techniques presented in the article may also help in automated identification and characterization of dynamical patterns arising in coupled dynamics on networks for various applications~\cite{Lee2018,Maksimenko2017}.

\section*{Acknowledgments}
SJ acknowledges DST project grant (EMR/2016/001921) and CSIR project grant (25 (0293)/18/ EMR - II) for financial support. SG acknowledges DST for the INSPIRE fellowship (IF150149).

\section*{Code and Sample Data Files availability:}
All codes used in the present article along with the sample files for the figures are publicly available on GitHub Repo \cite{githubdata}.

\end{document}